\documentclass[a4paper,12pt]{article}
\usepackage{jcappub}
\usepackage{amssymb}
\usepackage{graphicx}
\usepackage{amsmath}
\usepackage{hyperref}
\usepackage{tikz}
\usepackage{bm}
\usepackage{exscale}
\usepackage{relsize}
\usepackage{placeins}
\allowdisplaybreaks[4]

\usetikzlibrary{arrows.meta,decorations.pathmorphing,backgrounds,positioning,fit,petri,patterns}
\tikzset{zigzag/.style={decorate, decoration=zigzag}}

\definecolor{darkgreen}{rgb}{0.,0.6,0.}

\newcommand{\md}{\mathrm{d}}
\newcommand{\me}{\mathrm{e}}
\newcommand{\mD}{\mathcal{D}}

\title{\boldmath Cosmological Vacuum Decays from Schwinger-Keldysh Formalism}

\author[]{Zi-Yan Yuwen,}
\emailAdd{ziyan.yuwen@apctp.org}

\affiliation[]{Asia Pacific Center for Theoretical Physics (APCTP), Pohang 37673, Korea}

\abstract{
In this work, we take a first step to a stochastic real-time description of the vacuum phase transitions in a radiation-dominated FLRW universe from the Schwinger-Keldysh formalism. By splitting the phase transition field $\Phi$ into the long-wavelength mode $\phi$ and the short-wavelength mode $\sigma$ and tracing over the latter as the environment, we obtain a classical Langevin-type equation-of-motion for the long-wavelength field $\phi$, where the quantum effects are encoded in a non-Markov memory kernel and a non-Gaussian noise. As a phenomenological example, we start from a polynomial potential and a thermalized initial state, and study the structure of the memory kernel as well as the correlation functions of the noise term. Further extensions remain possible to describe the whole dynamics of cosmological first-order phase transitions via numerical simulations.  
}

\begin{document}
\maketitle
\flushbottom

\section{Introduction}\label{sec: introduction}

First-order phase transitions (FOPT) play an important role in cosmology~\cite{Hawking:1982ga,Witten:1984rs,Hogan:1986dsh} as a manifestation of nonequilibrium quantum field dynamics (see also~\cite{Mazumdar:2018dfl,Hindmarsh:2020hop,Caldwell:2022qsj,Athron:2023xlk} for recent reviews). They may occur in a variety of particle-physics settings in the early universe, giving rise to rich phenomenology including baryogenesis~\cite{Cohen:1990it,Cohen:1990py,Nelson:1991ab,Cohen:1993nk,Cohen:1994ss,Cohen:2012zza}, primordial black hole formation~\cite{Liu:2021svg,Hashino:2021qoq,Baker:2021nyl,Baker:2021sno,Kawana:2021tde,Jinno:2023vnr,Wang:2025hwc,Ning:2026nfs}, primordial magnetic fields~\cite{Hogan:1983zz,Quashnock:1988vs,Vachaspati:1991nm,Cheng:1994yr,Baym:1995fk}, \textit{etc.}. In an FOPT, the system initially trapped in a metastable vacuum tunnel toward the true vacuum, resulting in the nucleation of vacuum bubbles, followed by their subsequent expansion, percolation, and most importantly, gravitation wave radiations~\cite{Kamionkowski:1993fg,Jinno:2016vai,Hindmarsh:2016lnk,Hindmarsh:2019phv,Cai:2023guc,Guo:2020grp}.

Traditional methods on estimating the vacuum decay rates are mainly based on evaluating the Euclidean action of bounce solutions obtained via Wick rotation~\cite{Coleman:1977py,Callan:1977pt}, corresponding to a thermal equilibrium system with the temperature $T$ serving as the inverse of period $\beta$ along the time direction. While this approach was originally developed for vacuum decay in flat spacetime, it has subsequently been generalized to scenarios involving finite temperature effects~\cite{Linde:1980tt,Linde:1981zj}, cosmological constant~\cite{Coleman:1980aw,Hawking:1981fz}, primordial black holes~\cite{Moss:1984zf,Hiscock:1987hn,Gregory:2013hja,Burda:2015isa,Burda:2015yfa,Mukaida:2017bgd}, and primordial curvature perturbations~\cite{Yuwen:2026sub}. In contrast to this Euclidean approach, an alternative method estimates decay rates by examining the real-time dynamics of the quantum system through the Picard-Lefschetz theory~\cite{Tanizaki:2014xba,Ai:2019fri,Mou:2019gyl,Nishimura:2023dky}, where the contributions to the path integral arise from complex saddle points associated with Lefschetz thimbles. In a more realistic case, however, since the vacuum decays take place in an expanding universe via quantum tunnelling, a real-time nonequilibrium description is required to simultaneously account for quantum fluctuations and the effects of cosmological expansion. In particular, the effective dynamics of the order parameter (the field value) may receive non-trivial corrections from its self-interactions and coupling to to the evolving background geometry.

A natural framework for studying such questions is provided by the Schwinger-Keldysh (SK) formalism ~\cite{Schwinger:1960qe,Keldysh:1964ud}, also known as the closed-time-path formalism or in-in formalism~\cite{Weinberg:2005vy,Weinberg:2006ac}, which is designed for the real-time description of nonequilibrium quantum systems~\cite{Kamenev:2009jj,Polkovnikov:2009ys}. From the viewpoint of open quantum systems, one may separate the relevant coarse-grained degrees of freedom from the remaining modes and derive an effective description for the reduced system by tracing out the latter. In general, this procedure gives rise to non-local and stochastic dynamics: the environmental back-reaction appears in the form of dissipations, while quantum fluctuations are encoded in correlators of the random noise~\cite{Calzetta:2000tp,Calzetta:2001gv,Hertzberg:2019wgx}.  One of the most successful applications of open quantum system to cosmology is the stochastic inflation~\cite{Finelli:2008zg,LopezNacir:2011kk,PerreaultLevasseur:2013kfq,Moss:2016uix,Christodoulidis:2025vxz}. Such an approach is particularly also well suited to cosmological FOPT, where the long-wavelength dynamics of the transition field is influenced by ultraviolet or short-wavelength fluctuations in an intrinsically nonequilibrium setting. It also offers a systematic way to connect the underlying quantum field theory to an effective Langevin description beyond the simplest classical equation-of-motion~\cite{Yokoyama:2004pf,Miyamoto:2013gna,Kaplanek:2025moq,Kaplanek:2026kpp}. Recently, some work has applied the SK formalism to the studies on the non-equilibrium dynamics of the vacuum bubble wall during an FOPT~\cite{Ai:2025bjw}.
In this work, we formulate vacuum phase transition in a Friedmann-Lemaître-Robertson-Walker (FLRW) universe within this open-system framework based on the SK formalism. We decompose the phase-transition field into a long-wavelength field and short-wavelength modes, treat the latter as an environment, and derive an effective classical Langevin-type equation of motion for by tracing out the environmental degrees of freedom.

This paper is structured as follows. In Sec.~\ref{sec: action}, we present the construction of effective action for $\phi$ as an open system. In Sec.~\ref{sec: EOM}, we derive the Langevin-type equation-of-motion, and study the structure of memory kernel in \ref{subsec: memory kernel} and the noise term in \ref{subsec: noise} respectively. Our conclusion and discussions are summarized in Sec.~\ref{sec: conclusions}. Throughout this paper the metric signature follows the most-positive convention $(-,+,+,+)$, and the natural unit $c=G=\hbar=1$ is adopted.

\section{Effective Action}\label{sec: action}

\subsection{Full theory and scale split}

Our aim is to describe the vacuum phase transitions in a cosmological background. 
For simplicity, let us turn off the back-reaction of the perturbations to the spacetime, and fix the background as a radiation-dominated (RD) FLRW universe, since most scenarios of interest, such as electroweak and QCD phase transitions, occur well within the RD epoch. Consider an FOPT described by a real scalar field $\Phi$ with the following action,
\begin{align}
    S_0[\Phi] = \int \md \eta\md^3\bm{x} \> a^4\left( - \frac{1}{2}g^{\mu\nu} \partial_\mu \Phi \partial_\nu \Phi - \frac{1}{2}m^2 \Phi^2 - \frac{g}{3!}\Phi^3 - \frac{\lambda}{4!}\Phi^4  \right) ~,
\end{align}
where the background metric is given by $g_{\mu\nu} = a^2\mathrm{diag}(-1,1,1,1)$. There are temperature dependences of the effective mass $m$ and the three-point self-interaction parameters $g$~\cite{Quiros:1999jp},
\begin{align}\label{eq: msquare and g}
    m^2 = M^2 (T^2 - T_0^2)~, \quad g=-AT~,
\end{align}
with $A$ and $\lambda$ strictly positive. Although this result is derived from finite-temperature field theory, it remains valid as long as the cosmic expansion can be regarded as adiabatic, in which case local thermal equilibrium can be established. 

Next, we split the full field $\Phi$ as a summation of the long-wavelength mode $\phi$ and short-wavelength model $\sigma$ according to a characteristic scale $k_W$, 
\begin{align}
    \Phi(\eta,\bm{x}) = \phi(t,\bm{x}) + \sigma(t,\bm{x})~,
\end{align}
where the long-wavelength mode is obtained by applying a window function in the Fourier space that smooths out all modes with wavelengths shorter than the given scale $k_W$,
\begin{align}
    \phi(t,\bm{k}) = \Theta(1 - k/k_W) \, \Phi(t,\bm{k})~.
\end{align}
Then the full action can be re-written as
\begin{align}\label{eq: splitting S0}
    S_0[\Phi] = S_0[\phi] + S_0[\sigma] + S_\mathrm{int}[\phi, \sigma]~,
\end{align}
where the interaction reads
\begin{align}
    S_\mathrm{int}[\phi, \sigma] = \int \md \eta\md^3\bm{x} \> a^4 &\left[ \frac{\sigma' \phi' - \partial_i\sigma\partial_i \phi}{a^2}  - m^2 \sigma\phi \right. \nonumber\\
    &\quad\quad\quad
    \left.  - \left(\frac{g}{2}\sigma^2+\frac{\lambda}{6}\sigma^3\right)\phi
    -\left(\frac{g}{2}\sigma+\frac{\lambda}{4}\sigma^2\right)\phi^2
    -\frac{\lambda}{6}\sigma\phi^3
    \right]~\nonumber\\
    = \int \md \eta\md^3\bm{x} \> a^4 &\left[ - \left(\frac{\sigma'' + 2 \mathcal{H}\sigma' - \nabla^2\sigma}{a^2}+m^2\sigma\right)\phi   \right. \nonumber\\
    &\quad\quad\quad
    \left. - \left(\frac{g}{2}\sigma^2 +\frac{\lambda}{6}\sigma^3 \right)\phi
    -\left(\frac{g}{2}\sigma+\frac{\lambda}{4}\sigma^2\right)\phi^2
    -\frac{\lambda}{6}\sigma\phi^3\right]
    \label{eq: S_int}
\end{align}
where the prime denote the derivative with respect to the conformal time $\eta$, and $\nabla^2$ is the spatial Laplacian operator. The first term in Eq.~\eqref{eq: S_int} is obtained through integral by parts and subsequently neglecting the boundary terms, where $\mathcal{H}\equiv a'/a$ is the conformal Hubble parameter.

We focus on the evolution of the long-wavelength field $\phi$ as an open system, associating with the short-wavelength modes $\sigma$ as the environment where the fluctuations are encoded. The subsystem $\phi$ coarse-grained non-perturbative background component of $\Phi$, whereas the environment $\sigma$ corresponds to small fluctuations around $\Phi\simeq\phi$ and can be analysed with perturbation theory. 
We start with a finite initial time $\eta_i$ after reheating, and neglect connected correlations between the long- and short-wavelength modes at initial time as approximation. Then the initial density matrix takes a separable form $\hat\rho_{{\rm tot},i} = 
\hat\rho_{\phi,i} \otimes\hat\rho_{\sigma,i}$. The reduced density matrix for the open system $\phi$ is obtained by tracing over the environment $\sigma$, i.e. $\hat{\rho}(\eta) = \mathrm{Tr}_\sigma \hat{\rho}_\mathrm{tot} = \mathrm{Tr}_\sigma \left(\hat U(\eta,\eta_i) \hat{\rho}_{\mathrm{tot},i} \hat U^\dagger(\eta,\eta_i)\right)$.  
Consider a certain final state $|\phi_\mathrm{out}\rangle$, the diagonal term of the density matrix is given by two branches of path integrals,
\begin{align}
    \rho_{\phi_\mathrm{out}\phi_\mathrm{out}} = \langle\phi_\mathrm{out}|\hat{\rho}|\phi_\mathrm{out}\rangle = \int^{\phi_\mathrm{out}} \mD \phi_+ \mD\phi_- \,\hat{\rho}_{\phi,i}\,\me^{iS_\mathrm{eff}[\phi_+, \phi_-]}~,
\end{align}
where the effective action $S_\mathrm{eff}$ is usually referred to as the open functional~\cite{Colas:2025app}, which is obtained by tracing over $\sigma$,
\begin{align}
    \me^{iS_\mathrm{eff}[\phi_+, \phi_-]} &= \int \mD\sigma_+ \mD \sigma_- \,  \hat{\rho}_{\sigma,i} \,\me^{i(S_0[\Phi_+] - S_0[\Phi_-])} \nonumber\\
    &= \me^{i(S_0[\phi_+] - S_0[\phi_-])} \int \mD \sigma_+ \mD \sigma_- \, \hat{\rho}_{\sigma,i} \,\me^{i(S_0[\sigma_+] - S_0[\sigma_-] + \,S_\mathrm{int}[\phi_+, \sigma_+] - S_\mathrm{int}[\phi_-, \sigma_-])} \nonumber\\
    &\equiv \me^{i(S_0[\phi_+] - S_0[\phi_-] + F[\phi_+, \phi_-])}~,
\end{align}
where we have used Eq.~\eqref{eq: splitting S0}. It's clear to see that $S_\mathrm{eff}[\phi_+, \phi_-]$ consist of unitary dynamics of the system $S_0[\phi]$ and an influence functional describing effects from the environment $F[\phi_+, \phi_-]$~\cite{Feynman:1963fq}. To summarize, the evolution of $\phi$ can be described by a path integral in a closed-time contour, known as the Schwinger-Keldysh (SK) formalism~\cite{Schwinger:1960qe,Keldysh:1964ud} or in-in formalism~\cite{Weinberg:2005vy,Weinberg:2006ac}. It was shown that the in-in formalism is not only equivalent to the traditional in-out calculation in the case of pure state~\cite{Donath:2024utn}, but also accommodates mixed states upon tracing over the environment. The in-in formalism in a cosmological background provides a systematic perturbative framework for computing the correlation functions~\cite{Chen:2016nrs,Chen:2016uwp,Chen:2017ryl}, often referred to as the cosmological collider. In contrast, we study this framework in a RD FLRW background instead of an inflation background.

\subsection{Influence Functional}

Now we focus on the evaluation of influence functional $F[\phi_+,\phi_-]$, which can be computed by integrating over the environment order by order,
\begin{align}\label{eq: influence functional}
    F[\phi_+,\phi_-] = -i\ln\int\mD\sigma_+ \mD \sigma_- \, \hat{\rho}_\sigma \,\me^{i(S_0[\sigma_+] - S_0[\sigma_-])} \left( 1 + i\Delta S_\mathrm{int} - \frac{1}{2}\Delta S_\mathrm{int}^2 + \dots\right)
\end{align}
with abbreviation $\Delta S_\mathrm{int} \equiv S_\mathrm{int}[\phi_+, \sigma_+] - S_\mathrm{int}[\phi_-, \sigma_-]$, yielding
\begin{align}
    F[\phi_+,\phi_-] = \langle \Delta S_\mathrm{int} \rangle_\sigma + \frac{i}{2}\langle \Delta S_\mathrm{int}^2 \rangle_\sigma + \dots~,
\end{align}
where the bracket $\langle\dots\rangle_\sigma$ stands for tracing over $\sigma$. 

The results can be obtained by replacing internal $\sigma$ legs with the corresponding propagators. In order to get the Feynman rules, let us focus on the perturbation theory for $\sigma$ field. To begin with, the quadratic part of the action describing the free theory is given by
\begin{align}
    S^{(2)}_0[\sigma] = \int \md \eta \md^3\bm{x} \> a^4 \left(\frac{1}{2a^2}\sigma'^2 - \frac{1}{2a^2}\delta_{ij}\partial_i\sigma\partial_j\sigma - \frac{1}{2}m^2 \sigma^2 \right)~.
\end{align}
The spatial Fourier transform of $\sigma$ on a given equal-time slice is defined as
\begin{align}
    \sigma(\eta,\bm{x}) = \int \frac{\md^3\bm{k}}{(2\pi)^3}\>\sigma_{\bm{k}}(\eta) \me^{i\bm{k}\cdot\bm{x}}~.
\end{align}
Then following the standard  canonical quantisation procedure, the field variables are promoted to operators on the Hilbert space,
\begin{align}
    \hat{\sigma}_{\bm{k}}(\eta) = \sigma_{{k}}(\eta) \hat{a}_{\bm{k}} + \sigma_{{k}}^*(\eta) \hat{a}_{\bm{k}}^\dagger~,
\end{align}
where $\hat{a}_{\bm{k}}$ and $\hat{a}_{\bm{k}}^\dagger$ are the annihilation and creation operators defined with respect to the vacuum $|\Omega\rangle_\sigma$. The mode function $\sigma_{{k}}(\eta)$ satisfies the free equation-of-motion (EOM) in Fourier space,
\begin{align}\label{eq: EOM of sigma}
    \sigma_{{k}}'' + 2\mathcal{H}\sigma_{{k}}' + (k^2 + a^2m^2) \sigma_{{k}} = 0~.
\end{align}
One can scale the field variable by introducing an auxiliary variable $\chi = a \sigma$, whose EOM is given by the Mukhanov-Sasaki Equation,
\begin{align}
    \chi_{{k}}'' + \left(k^2 + a^2m^2 -\frac{a''}{a} \right) \chi_{{k}} = 0~.
    \label{eq: EOM of chi}
\end{align}
The effective mass is given by
\begin{align}
    \tilde{m}^2 &= a^2m^2 - \frac{a''}{a} = M^2T_0^2 \left(1-\frac{a^2}{a_0^2} \right) - \frac{a''}{a} \nonumber\\
    &= m_0^2 \left(1 - \frac{\eta^2}{\eta_0^2} \right) - \frac{2(1-3w)}{(1+3w)^2}\,\eta^{-2}~,
    \label{eq: mode function chi}
\end{align}
where $m_0\equiv MT_0$, and $w$ is the EoS parameter of the background fluid.
In our specific case, the background is chosen to be RD FLRW satisfying $w=1/3$, and $a''=0$, and thereby the last term vanishes. 
Taking an adiabatically expansion approximation, the entropy per comoving spatial volume is approximately a constant $S/V_\mathrm{com}= \frac{2\pi^2}{45} g_{*,s} T^3a^3 = \mathrm{const.}$, thereby here we adopt $T\propto1/a$ by assuming constant entropy degrees of freedom. $m_0$ can be interpreted as the mass at initial time $\eta_i\ll \eta_0$. The effective mass becomes negative after the critical time $\eta_0$ and leads to tachyonic instability at late time. Solving the EOM for each $\bm{k}$ and using the Wronskian conditions gives the mode function as follows
\begin{align}\label{eq: full mode function}
    \chi_{ k}(\eta) = \left(\frac{\Gamma(\mu)}{\sqrt{4\pi\gamma}}\right)^{1/2} D_{-\mu}\left(i\sqrt{2\gamma}\,\eta \right)~,
\end{align}
where $D_\alpha$ is the Parabolic Cylinder $D$ function, $\mu=(k^2+m_0^2)/(2\gamma) + 1/2$, and $\gamma = m_0/\eta_0$. For an early time $\eta\ll \eta_0$, the mode function can be approximated as a plane wave solution  with dispersion relation $E_k^2= k^2+m_0^2$, 
\begin{align}\label{eq: chi plane wave}
    \chi_{ k}(\eta) \simeq \frac{1}{\sqrt{2E_k}} \me^{-iE_k \eta}~.
\end{align}
As the temperature reduces close to the critical temperature $T_0$, the mode function approaches the massless limit $\tilde{m}^2\to 0$. If the system supercools below $T_0$, the true vacuum becomes the only local minimum, the false vacuum becomes spinodally unstable and the relevant dynamics changes from barrier crossing to tachyonic growth with $\tilde{m}^2<0$. Such a regime requires non-perturbative approach to study the tachyonic instability and lies beyond the present false-vacuum-decay analysis.

Then let us consider the self-interactions. It is easy to see that the interaction vertices reduce to similar forms of the flat case which has no time dependence, by absorbing the scale factors into the coupling parameters $\tilde{g} = ag$. Re-writing $S_0[\sigma]$ in terms of $\sigma$ and the new coupling constants $\tilde{m}$ and $\tilde{g}$ leads to the following effective action of $\chi$,
\begin{align}\label{eq: action of chi}
    S[\chi] &= S_0[\sigma = \chi /a] \nonumber\\
    &=\int\md\eta\md^3\bm{x} \left(\frac{1}{2}({\chi'}^2 - (\partial_i\chi)^2) - \frac{1}{2}\tilde{m}^2\chi^2 - \frac{1}{6}\tilde{g}\chi^3 - \frac{1}{24}\lambda \chi^4\right)~,
\end{align}
where in the second line we cancelled out the terms evolving scale factors by performing a integral by parts and neglecting the boundary terms as follows,
\begin{align}
\int\md\eta \left(-2\mathcal{H}\chi\chi' + (\mathcal{H}\chi)^2 \right) = \mathrm{Boundary~terms} + \int \md\eta\left( \mathcal{H}'\chi^2 + \mathcal{H}^2\chi^2 \right) = 0~.
\end{align}
The exact cancellation $\mathcal{H}'+\mathcal{H}^2=0$  is true only for RD FLRW background. For a generic FLRW background, the failing of cancellation actually leads to the $a''/a$ correction in the dispersion relation in Eq.~\eqref{eq: mode function chi}. 

It is clear to see that the theory for $\chi$ reproduces a scalar field in a flat background by replacing the time $t$ with the conformal time $\eta$. Therefore, the Feynman rules for $\sigma$ in a RD FLRW universe can be translated to those for $\chi$ living in a Minkowski spacetime. Since in the influence functional there is a double-copy of $\sigma$, each vertex can either be a ``$+$'' or ``$-$'' vertex connected by four types of internal lines. The expressions of propagators also depend on the choice of initial state for the environment. For an initial vacuum state, the forward propagators of internal lines are given by
\begin{align}\label{eq: G> vacuum}
    G^{-+}({k}; \eta_1, \eta_2) \equiv G^{>}({k}; \eta_1, \eta_2) = \sigma_{{k}}(\eta_1)\sigma^*_{{k}}(\eta_2) = \frac{1}{a(\eta_1)a(\eta_2)}\chi_{{k}} (\eta_1)\chi^*_{{k}}(\eta_2) ~.
\end{align}
Instead, if the long-wavelength mode is still described by a coarse-grained order parameter whose initial density matrix is localized near the high-temperature minimum $\phi=0$, while the short-wavelength modes are assumed to have approximately thermalized with the radiation plasma at physical temperature $T_i$, the propagators are modified as~\cite{Ghiglieri:2020dpq}
\begin{align}\label{eq: G> thermal}
    G^{-+}({k}; \eta_1, \eta_2) = (1+n_k) \sigma_{{k}}(\eta_1)\sigma^*_{{k}}(\eta_2) + n_k \sigma^*_{{k}}(\eta_1)\sigma_{{k}}(\eta_2)~,
\end{align}
where $n_k$ is the particle occupation number of the initial state for the modes with norm $|\bm{k}|=k$, satisfying the Bose-Einstein distribution 
\begin{align}
    n_k = \frac{1} {\me^{E_k/(a_iT_i)} - 1} ~.
\end{align}
Since at initial time $\eta_i$ the mode functions can be well approximated as plane waves, the energy at mode $k$ can be replaced by $E_k^2 = k^2 + m_0^2$.
In the limit $E_k\gg a_iT_i$, the initial thermal bath is unable to efficiently excite these short-wavelength modes, $n_k$ becomes negligible and the propagator expression Eq.~\eqref{eq: G> thermal} reduces to its vacuum version Eq.~\eqref{eq: G> vacuum}. 
Then the backward and (anti-)time-ordered propagators of internal lines are given by
\begin{align}
    G^{+-}({k}; \eta_1, \eta_2) &= G^{-+}({k}; \eta_1, \eta_2)^* \equiv G^{<}({k}; \eta_1, \eta_2) \\
    G^{\pm\pm}({k}; \eta_1, \eta_2) &= G^{\mp\pm}({k}; \eta_1, \eta_2)\Theta(\eta_1-\eta_2) + G^{\pm\mp}({k}; \eta_1, \eta_2)\Theta(\eta_2-\eta_1)~,
\end{align}
where $\Theta(x)$ is the Heaviside Theta function. Leveraging the invariance by spatial translation (homogeneity), the equal-time connected correlation functions is obtained by the spatial Fourier transform,
\begin{align}\label{eq: Gab of sigma}
    G^{\mathsf{ab}}(\eta_1,\bm{x}_1;\eta_2,\bm{x}_2) = G^{\mathsf{ab}}(\eta_1,\eta_2;\Delta\bm{x}) = \int \frac{\md^3\bm{k}}{(2\pi)^3} G^{\mathsf{ab}}({k}; \eta_1, \eta_2) \, \me^{i\bm{k}\cdot\Delta\bm{x}}
\end{align}
with $\mathsf{a},\,\mathsf{b} = +,\,-$ running over four types of propagators.

Now we can integrate out the $\sigma$ field by the propagators above. The linear-order $\langle \Delta S_\mathrm{int} \rangle_\sigma$ vanishes because of the tadpole removing by normal ordering of composite operators $O(\sigma)$ in the initial density matrix of the $\sigma$ field~\cite{Boyanovsky:2015xoa}, $O(\sigma)\to O(\sigma) - \langle O(\sigma)\rangle$. Therefore, the leading-order contribution comes from the second-second order expansion. Let us define the following currents and operators for later conveniences,
\begin{align}
\begin{aligned}
    J_n(\phi) &= \phi^n~,\quad n=1,2,3, \\
    O_1(\sigma) &= \frac{\sigma'' + 2 \mathcal{H}\sigma' - \nabla^2 \sigma}{a^2}+m^2\sigma + \frac{g}{2}\sigma^2 +\frac{\lambda}{6}\sigma^3~, \\
    O_2(\sigma) &= \frac{g}{2}\sigma + \frac{\lambda}{4}\sigma^2 ~,
    \quad O_3(\sigma) = \frac{\lambda}{6}\sigma~. 
\end{aligned}
\end{align}
The connected second-order contribution is expressed in terms of $J_n$ and expectation values of $O_n$,
\begin{align}\label{eq: F2 Jnm}
    F = \frac{i}{2} \sum_{n,m=1}^3 \sum_{\mathsf{a,b}=\pm} \mathsf{ab} \iint_{x,y} J_n(\phi_\mathsf{a}(x)) G_{nm}^{\mathsf{ab}}(x,y) J_m(\phi_\mathsf{b}(y))~,
\end{align}
where $x=(\eta_1,\bm{x})$ and $y=(\eta_2,\bm{y})$ for short, and the integral volume element is defined as $\int_x\equiv\int a^4\md\eta\md^3\bm{x}$. The connected two-point correlators are defined as
\begin{align}
\begin{aligned}
    & G_{nm}^{-+}(x,y) = \big\langle O_n(\sigma_\mathsf{a}(x))  O_m(\sigma_\mathsf{b}(y)) \big\rangle \equiv G_{nm}^{>}(x,y) ~, \\ 
    & G_{nm}^{+-}(x,y) = \big\langle O_m(\sigma_\mathsf{b}(y)) O_n(\sigma_\mathsf{a}(x)) \big\rangle = G_{nm}^{-+}(x,y)^* \equiv G_{nm}^{<}(x,y)~,\\
    & G_{nm}^{++}(x,y) = \big\langle T O_n(\sigma_\mathsf{a}(x)) O_m(\sigma_\mathsf{b}(y)) \big\rangle~, \quad G_{nm}^{+-}(x,y) = \big\langle \tilde{T} O_n(\sigma_\mathsf{a}(x)) O_m(\sigma_\mathsf{b}(y)) \big\rangle~,
\end{aligned} 
\end{align}
where $T$ and $\tilde{T}$ stand for time ordering and anti-time ordering, respectively. These correlators of composite operators can be expressed in terms of $G^{\mathsf{ab}}$ defined in \eqref{eq: Gab of sigma} through Wick expansion.
One can reorganize the expressions in the Keldysh basis by a linear combination of $+$ and $-$ fields on the two branches of the closed-time contours~\cite{Keldysh:1964ud,Kamenev_2011},
\begin{align}
    \phi_c=\frac{\phi_+ + \phi_-}{2}~,\quad \phi_\Delta=\phi_+ - \phi_- ~.
\end{align}
In other literatures (e.g.~\cite{Colas:2025app}) they are also named as the retarded $\phi_r$ and advanced components $\phi_a$, respectively. Here we use the subscript $c$ and $\Delta$ instead to denote for the on-shell classical component and the quantum perturbations deviating from the classical EOM~\cite{Hertzberg:2019wgx,Alvestad:2025iep}. Same Keldysh rotations can be applied to $\sigma$ and composite operators $J$, 
\begin{align}
    &J_1^c = \phi_c~,&& J_2^c = \phi_c^2 + \frac{1}{4}\phi_\Delta^2~,&& J_3^c = \phi_c^3 + \frac{3}{8} \phi_c\phi_\Delta^2~, \\
    &J_1^\Delta = \phi_\Delta~,&& J_2^\Delta = 2\phi_c\phi_\Delta~,&& J_3^\Delta = 3\phi_c^2\phi_\Delta + \frac{1}{4} \phi_\Delta^3~.
\end{align}
The summations over $\mathsf{a}$ and $\mathsf{b}$ in Eq.~\eqref{eq: F2 Jnm} then is expressed as
\begin{align}
    \sum_{\mathsf{a,b}=\pm} &\mathsf{ab} \> J_{n,a} G_{nm}^{\mathsf{ab}}J_{m,b} = [J_n^+, J_n^-] \left[\begin{array}{cc}
        G_{nm}^{++} & -G_{nm}^{+-} \\
        -G_{nm}^{-+} & G_{nm}^{--}
    \end{array} \right]
    \left[\begin{array}{c}
        J_m^+ \\ J_m^-
    \end{array} \right] = [J_n^c, J_n^\Delta]
    \left[\begin{array}{cc}
        0 & D_{nm}^{c\Delta} \\
        D_{nm}^{\Delta c} & N_{nm}
    \end{array} \right]
    \left[\begin{array}{c}
        J_m^c \\ J_m^\Delta
    \end{array} \right]~,
\end{align}
where we have used the fact that 
\begin{align}
    G_{nm}^{++}(x,y) + G_{nm}^{--}(x,y) = G_{nm}^{-+}(x,y) + G_{nm}^{+-}(x,y)~.
\end{align}
The linear combinations of correlation functions are evaluated as
\begin{align}
    D_{nm}^{\Delta c} &= \frac{1}{2}\left(G_{nm}^{++} + G_{nm}^{-+}  - G_{nm}^{--} - G_{nm}^{+-}\right) \nonumber\\
    &= \Theta(\eta_1-\eta_2) (G_{nm}^{>}  - G_{nm}^{<}) = \Theta(\eta_1-\eta_2)\langle[O_n,O_m]\rangle \equiv \Theta(\eta_1-\eta_2)  \tilde{D}_{nm}^{\Delta c} ~, \\
    D_{nm}^{c \Delta} &= \frac{1}{2}\left(G_{nm}^{++} + G_{nm}^{+-} - G_{nm}^{--} - G_{nm}^{+-}\right) = -\Theta(\eta_2-\eta_1) \langle[O_n,O_m]\rangle = D_{mn}^{\Delta c}(x\leftrightarrow y)~, \\
    N_{nm} &= \frac{1}{4}\left(G_{nm}^{++} + G_{nm}^{-+}  + G_{nm}^{--} + G_{nm}^{+-}\right) = \frac{1}{2}(G_{nm}^{>} + G_{nm}^{<}) = \frac{1}{2} \langle\{O_n,O_m\}\rangle~.
\end{align}
By defining $D_{nm} =-iD_{nm}^{\Delta c} = -i\Theta(\eta_1-\eta_2) \langle[O_n,O_m]\rangle$, finally the influence functional to the second-order is expressed in a a compact matrix multiplication form,
\begin{align}\label{eq: F matrix form}
    F= -\iint_{x,y} \mathbf{J}_\Delta ^\top(x) \,\mathbf{D}(x,y) \, \mathbf{J}_c(y) + \frac{i}{2} \iint_{x,y}\mathbf{J}_\Delta^\top(x) \,\mathbf{N}(x,y)\,\mathbf{J}_\Delta(y)~,
\end{align}
where $\mathbf{J}_{\Delta/c}$ is the current vector $[J_n^{\Delta/c}]$, with kernel matrices $\mathbf{D} = [D_{nm}]$ and $\mathbf{N}=[N_{nm}]$, for $n,m=1,2,3$.
Here we list all the components of $\mathbf{D}$ explicitly,
\begin{subequations}\label{eq: Dnm components}
\begin{align}
    \tilde{D}_{11}^{\Delta c} =&\> \frac{\partial_1^2 + 2\mathcal{H}_1\partial_{\eta_1} + \tilde{m}^2}{a_1^2} \, \frac{\partial_2^2 + 2\mathcal{H}_2\partial_2 + \tilde{m}^2}{a_2^2} \big(G^>(x,y) - G^<(x,y)\big)  \nonumber\\
    &\> + \frac{\tilde{g}^2}{2a_1 a_2} \big(G^>(x,y)^2 - G^<(x,y)^2\big) + \frac{\lambda^2}{6} \big(G^>(x,y)^3 - G^<(x,y)^3\big) ~,\\
    \tilde{D}_{22}^{\Delta c} =&\> \frac{\tilde{g}^2}{4a_1 a_2} \big(G^>(x,y) - G^<(x,y)\big) + \frac{\lambda^2}{8}  \big(G^>(x,y)^2 - G^<(x,y)^2\big)~, \\
    \tilde{D}_{33}^{\Delta c} =&\> \frac{\lambda^2}{36}  \big(G^>(x,y) - G^<(x,y)\big)~, \\
    \tilde{D}_{12}^{\Delta c} =&\> \frac{\partial_1^2 + 2\mathcal{H}_1\partial_{\eta_1} + \tilde{m}^2}{a_1^2}\frac{\tilde{g}}{2a_2}\big(G^>(x,y) - G^<(x,y)\big) +\frac{\tilde{g}\lambda}{4a_1}\big(G^>(x,y)^2 - G^<(x,y)^2\big)~,\\
    \tilde{D}_{13}^{\Delta c} =&\> \frac{\partial_1^2 + 2\mathcal{H}_1\partial_{\eta_1} + \tilde{m}^2}{a_1^2}\frac{\lambda}{6}\big(G^>(x,y) - G^<(x,y)\big)~, \\
    \tilde{D}_{23}^{\Delta c} =&\> \frac{\tilde{g}\lambda}{12a_1}  \big(G^>(x,y) - G^<(x,y)\big)~,
\end{align}
\end{subequations}
where we have used the abbreviations $\partial_A^2\equiv\partial_{\eta_{A}}^2-\nabla^2$, $a_A\equiv a(\eta_A)$, $\mathcal{H}_A\equiv \mathcal{H}(\eta_A)$ for $A=1,2$, and the local contact/tadpole terms have been removed from the non-local propagators by normal ordering. The components of $\mathbf{N}$ can be easily obtained by replacing the commutators with anti-commutators in the equations above. As is shown in Eq.~\eqref{eq: Dnm components}, the central structures in the kernels are the linear combinations of powers of propagators $G^>(x,y)^n \pm G^<(x,y)^n = G^>(x,y)^n \pm G^>(x,y)^{n,*}$. Therefore, for a generic complex propagator $G^>(x,y)$, the two kernels correspond to the imaginary and real parts respectively,
\begin{align}
    D_{nm} = \Theta(\eta_1-\eta_2) \> \mathrm{Im} \, G_{nm}^{>}~,\quad N_{nm}=\frac{1}{2}\mathrm{Re} \, G_{nm}^{>}~.
\end{align}

\section{Langevin-type Equation of Motion}\label{sec: EOM}

With the influence functional at hand, effective action now reads
\begin{align}\label{eq: Seff}
    S_\mathrm{eff} =S_0[\phi_+]-S_0[\phi_-]-\iint_{x,y} \mathbf{J}_\Delta ^\top(x) \,\mathbf{D}(x,y) \, \mathbf{J}_c(y) + \frac{i}{2} \iint_{x,y}\mathbf{J}_\Delta^\top(x) \,\mathbf{N}(x,y)\,\mathbf{J}_\Delta(y)~.
\end{align}
Taking the variation with respect to $\phi_\Delta$, one arrives at the effective EOM for $\phi_c$. In addition to the classical dynamics, the EOM contains memory corrections and stochastic noises.
The first tow terms in Eq.~\eqref{eq: Seff} are purely real, basically corresponding to the classical EOM, following by a purely real memory term, while the last integral is purely imaginary in a quadratic form of $J_\Delta$. For later convection, we split the current $\mathbf{J}_{c/\Delta}$ into the linear terms and non-linear terms,
\begin{align}
    \mathbf{J}_c &= \mathbf{J}_c^\mathrm{lin} + \mathbf{J}_c^\mathrm{nl}\phi_\Delta^2~, && \mathbf{J}_c^\mathrm{lin} = \left[\phi_c,\phi_c^2,\phi_c^3\right]^\top~, && \mathbf{J}_c^\mathrm{nl}= \left[0,1/4,3\phi_c/8\right]^\top~, \\
    \mathbf{J}_\Delta &= \mathbf{J}_\Delta^\mathrm{lin}\phi_\Delta + \mathbf{J}_\Delta^\mathrm{nl}\phi_\Delta^3~, && \mathbf{J}_\Delta^\mathrm{lin} = \left[1,2\phi_c,3\phi_c^2\right]^\top~, && \mathbf{J}_c^\mathrm{nl}= \left[0,0,1/4\right]^\top ~.
\end{align}
Then the effective action  in the Keldysh basis with all the terms sorted in the order of $\phi_\Delta$ is given by
\begin{align}
    S_\mathrm{eff} =&\> -\int_{x} \bigg(\frac{\phi_c''(x) + 2\mathcal{H}\phi_c'(x) - \nabla^2\phi_c(x)}{a^2}+m^2\phi_c(x) + \frac{g}{2}\phi_c^2(x)+\frac{\lambda}{6}\phi_c^3(x) \nonumber\\
    &\> \quad\quad + \mathbf{J}_\Delta^{\mathrm{lin}}(x)^\top \int_{y}  {\mathbf{D}}(x,y) \mathbf{J}_c^\mathrm{lin}(y) \bigg) \phi_\Delta(x)\nonumber\\
    &\>+ \frac{i}{2}\iint_{x,y} \mathbf{J}_\Delta^{\mathrm{lin}}(x)^\top {\mathbf{N}}(x,y)\mathbf{J}_\Delta^{\mathrm{lin}}(y) \phi_\Delta(x)\phi_\Delta(y)
    \nonumber\\
    &\> - \iint_{x,y} \mathbf{J}_\Delta^{\mathrm{lin}}(x)^\top {\mathbf{D}}(x,y)\mathbf{J}_c^{\mathrm{nl}}(y) \phi_\Delta(y)^2 \nonumber\\
    &\> - \int_{x} \left(\frac{g+\lambda\phi_c(x)}{24} + \mathbf{J}_\Delta^\mathrm{nl}(x)^\top \int_{y}{\mathbf{D}}(x,y) \mathbf{J}_c^\mathrm{lin}(y) \right)\phi_\Delta^3(x) \nonumber\\
    &\> + i \iint_{x,y} \mathbf{J}_\Delta^{\mathrm{lin}}(x)^\top {\mathbf{N}}(x,y)\mathbf{J}_\Delta^{\mathrm{nl}}(y) \phi_\Delta(x)\phi_\Delta^3(y) \nonumber\\
    &\> - \iint_{x,y} \mathbf{J}_\Delta^{\mathrm{nl}}(x)^\top {\mathbf{D}}(x,y)\mathbf{J}_c^{\mathrm{nl}}(y) \phi_\Delta^3(x)\phi_\Delta^2(y)  \nonumber \\
    &\> + \frac{i}{2} \iint_{x,y} \mathbf{J}_\Delta^{\mathrm{nl}}(x)^\top {\mathbf{N}}(x,y)\mathbf{J}_\Delta^{\mathrm{nl}}(y) \phi_\Delta^3(x)\phi_\Delta^3(y)~,
    \label{eq: Seff final}
\end{align}
where the spatial gradience of $\phi_c$ has been neglected. After variation, the terms that are linearly proportional to $\phi_\Delta$ simply indicates the dynamics involving only $\phi_c$. For the terms that are in a quadratic form of $\phi_\Delta$, one can perform a mathematical trick named Hubbard-Stratonovich (HS) transformation~\cite{Stratonovich:1957,Hubbard:1959ub} to rewrite the term as a bilinear function of $\phi_\Delta$ and a Gaussian random field $\xi$,
\begin{align}
    &\exp\left(-\frac{1}{2}\iint_{x,y}\phi_\Delta(x) K(x,y)\phi_\Delta(y)\right) \nonumber\\
    =&\> \sqrt{|\det K|} \int\mD\xi \, \exp\left(-\frac{1}{2}\iint_{x,y}\xi(x) K^{-1}(x,y)\xi(y) + i\int_x \xi(x)\phi_\Delta(x)\right)~,
\end{align} 
with a two point-correlation $\langle\xi(x)\xi(y)\rangle = K(x,y)$.
However, for the terms in cubic form of $\phi_\Delta$, a more complicated HS transformation is needed~\cite{Chakrabarty:2019aeu,Salcedo:2024smn} by introducing a random field $\xi$ with non-Gaussinity. 

Here we provide a generalized HS transformation to map all higher-order $\phi_\Delta$ terms to linear-order by introducing a random field $\xi$.
First of all, it has already been proved~\cite{Colas:2025app} that all odd powers of $\phi_\Delta$ are purely real and even powers of $\phi_\Delta$ purely imaginary, which suggests us to write down all the higher-order $\phi_\Delta$ contributions in $S_\mathrm{eff}$ in the following polynomial form,
\begin{align}
    i S_\mathrm{eff}\supset&\> - \frac{1}{2}\iint_{x_1,x_2} K_2(x_1,x_2)\phi_\Delta(x_1)\phi_\Delta(x_2) \nonumber \\
    &\> -\frac{i}{3!}\iiint_{x_1,x_2,x_3} K_3(x_1,x_2,x_3)\phi_\Delta(x_1)\phi_\Delta(x_2)\phi_\Delta(x_3) + \dots \nonumber\\
    =&\> \sum_{n=2} \left(\prod_{m=1}^n\int_{x_m}i\phi_\Delta(x_m)\right) \frac{K_n(x_1,\dots,x_n)}{n!} \equiv iS_\mathrm{eff}^\Delta~.
\end{align}
Because of the invariance under permutation of the integral variables, the integrals are symmetric under the exchange of any two arguments, and therefore one can define $K_n$ in a symmetric form
\begin{align}
    K_n(x_1,\dots,x_n) = \frac{1}{n!}\sum_{\pi\in S_n} K_n(x_{\pi(1)},\dots,x_{\pi(1)})~,
\end{align}
where $S_n$ is the permutation group of degree $n$. We can introduce a random field $\xi$ with probability distribution function (PDF)  $\mathcal{P}(\xi)$, such that $S_\mathrm{eff}^\Delta$ can be related the generating functional of $\xi$ with the external source given by $\phi_\Delta$,
\begin{align}
    Z[\phi_\Delta] = \mathcal{N} \int\mD\xi \> \mathcal{P}(\xi) \exp\left( i\int_x \xi(x) \phi_\Delta(x)\right)~,
\end{align}
where $\mathcal{N}$ is a normalization constant. More precisely, we choose a proper $\mathcal{P}(\xi)$ to satisfy the following condition, $\exp(iS_\mathrm{eff}^\Delta) = Z[\phi_\Delta]$. This can be achieved by a full description of the statistical properties of the random field. Specifically, the PDF is required to give the correct correlation functions to all orders, which is obtained by taking variational derivatives of the generating functional with respect to the source,
\begin{align}
    C_n\equiv \langle \xi(x_1)\dots \xi(x_n) \rangle = (-i)^n\frac{\delta}{\delta\phi_\Delta(x_1)} \dots \frac{\delta}{\delta\phi_\Delta(x_n)} Z[\phi_\Delta] \bigg|_{\phi_\Delta=0}~.
\end{align}
By identifying $\exp(iS_\mathrm{eff}^\Delta)$ and $Z[\phi_\Delta]$, one can calculate the correlation functions to all orders. For example, the correlation functions up to fifth order are given by
\begin{align}
\begin{aligned}
    C_1 &= 0~, \quad C_2 = K_2(x_1,x_2)~, \quad C_3 = K_3(x_1,x_2,x_3)~, \\
    C_4 &= K_4(x_1,\dots,x_4) + \left(K_2(x_1,x_2)K_2(x_3,x_4) + 2\mathrm{~more~permutations}\right)~, \\
    C_5 &= K_5(x_1,\dots,x_5) + \left(K_2(x_1,x_2)K_3(x_3,x_4,x_5) + 9\mathrm{~more~permutations}\right)~.
\end{aligned}
\label{eq: Ci def}
\end{align}
As long as all $C_n$ are matched, in principle the PDF can be reconstructed (but not necessarily to be unique). In practice, we don't have to write down an explicit form of $\mathcal{P}(\xi)$, as what we're interested in is actually the statistical properties of $\xi$ rather than the PDF itself.

Finally, the effective EOM takes the form of a Langevin-type stochastic differential equation,
\begin{align}\label{eq: Langevin EOM}
\begin{aligned} 
    \phi_c''(x)+2\mathcal{H}\phi_c'(x) - \nabla^2\phi_c(x) + \tilde{m}^2\phi_c(x) + \frac{a\tilde{g}}{2}\phi_c^2(x) + \frac{a^2\lambda}{6}\phi_c^3(x) & \\
    + a^2 \mathbf{J}_\Delta^{\mathrm{lin}}(x)^\top \int_y \, {\mathbf{D}}(x,y) \, \mathbf{J}_c^\mathrm{lin}(y) &= a^2{\xi}(x)~,
\end{aligned}
\end{align}
The integral kernel $\mathbf{D}$ proportional to $\Theta(\eta_1-\eta_2)$ characterizes how the states in the past affects the motion at the current time, and is therefore named as the ``memory kernel''. 
A random field $\xi(x)$ appearing on the right-hand-side of the EOM corresponds to the stochastic noise induced by the non-linear interactions between the long- and short-wavelength sectors, whose statistical properties are totally encoded in the propagators of the short-wavelength modes. In principle the noise term has non-trivial higher-order correlations, while to the leading order it may be approximated as a Gaussian random field. 

The EOM is a non-local integro-differential equation with a random noise, but can be simplified if we take a further step to consider the small $k_W$ limit, where almost only the zero-mode left in $\phi$.
Using the spatial translation invariance, the integral with respect to $\bm{y}$ results in an averaged memory kernel defined as
\begin{align}\label{eq: D bar}
    \bar{\mathbf{D}}(\eta_1,\eta_2) \equiv \int \md^3\bm{r} \, \mathbf{D}(\eta_1,\bm{r};\eta_2,\bm{0})~.
\end{align}
We can further re-define averaged variables by integrating $\bm{x}$ over a spatial volume $\mathcal{V}$. The $\phi_c$ field is not affected as it does not depend on space in this limit, i.e. $\bar{\phi}_c = \phi_c(\eta)$, while the noise now reads
\begin{align}\label{eq: xi bar}
    \bar{\xi}(\eta) = \int_\mathcal{V} \md^3\bm{x} \, \xi(\eta, \bm{x}) 
\end{align}
By averaging the EOM spatially, one arrives at
\begin{align}
    \phi_c''(\eta)+2\mathcal{H}\phi_c'(\eta) + \tilde{m}^2\phi_c(\eta) + \frac{a\tilde{g}}{2}\phi_c^2(\eta) + \frac{a^2\lambda}{6}\phi_c^3(\eta) & \nonumber \\
    + a^2 \mathbf{J}_\Delta^{\mathrm{lin}}(\eta)^\top \int_0^{\eta}\md\eta' \, a(\eta')^4 \, \bar{\mathbf{D}}(\eta,\eta') \, \mathbf{J}_c^\mathrm{lin}(\eta') &= a^2\bar{\xi}(\eta)~,
\end{align}
Before continue on the calculation, one thing should be reminded that the spatial integrals in \eqref{eq: D bar} and \eqref{eq: xi bar} have different physical origins, and therefore should be treated separately. The spatial average on the memory kernel is originated from the effective action and then should be integrated over the whole space. The spatial average on the noise term depends on the length scale $L\sim \mathcal{V}^{1/3}$ we are interested in. A larger $L$ means a coarser coarse-graining of the system and a smaller contribution from noise. We will return to this point quantitatively in Sec.~\ref{subsec: noise}.

\subsection{Memory kernel}\label{subsec: memory kernel}

The memory kernels $\bar{\mathbf{D}}$ requires spatial integral over the structure $G^{>,n} - G^{<,n}$ (or to say $\mathrm{Im}\,G^{>,n}$, which has a clear and intuitive correspondence to the Feynman diagrams. Here we calculate the result up to the $1$-loop level.

\subsubsection{Tree level}
To the tree level, the central structure is given by $G^{>} - G^{<}$, which can be found in every component of $\bar{\mathbf{D}}$, corresponding to the internal lines $\sigma$ shown below
\begin{align}
    \parbox{30mm}{\centering
    \begin{tikzpicture}
        \draw (-1,0) node[below] {$x$} circle (0.1);
        \fill (1,0)  node[below] {$y$} circle (0.1);
        \draw (-0.9,0) -- (0,0) node[above] {$\sigma$} -- (1,0);
    \end{tikzpicture}
    } ~-~
    \parbox{30mm}{\centering
    \begin{tikzpicture}
        \fill (-1,0) node[below] {$x$} circle (0.1);
        \draw (1,0)  node[below] {$y$} circle (0.1);
        \draw (-1,0) -- (0,0) node[above] {$\sigma$} -- (0.9,0);
    \end{tikzpicture}
    }
    = G^>(x,y)-G^<(x,y)~,
\end{align}
where the white dot denotes for a $-$ node and black dot denotes for a $+$ node. The vertex on each side can connect to $1$ to $3$ external lines $\phi$, contributing to pre-factors (pre-operators) listed as follows
\begin{align}\label{eq: phi-sigma}
    \parbox{25mm}{
    \begin{tikzpicture}
        \draw (0,0) node[below] {$x$} circle (0.1);
        \fill[pattern=north east lines] (0,0) circle (0.1);
        \draw (-1,0) node[left] {$~$} --  (-0.5,0) node[above] {$\phi$}  -- (-0.1,0);
        \draw (0.1,0) -- (0.5,0) node[above] {$\sigma$} -- (1,0);
    \end{tikzpicture} 
    } 
    &\quad\to\quad \int_x \frac{\partial_\eta^2 + 2 \mathcal{H}\partial_\eta - \nabla^2 + \tilde{m}^2}{a^2} \, (...) \\
    \parbox{25mm}{
    \begin{tikzpicture}
        \draw (0,0) node[below] {$x$} circle (0.1);
        \fill[pattern=north east lines] (0,0) circle (0.1);
        \draw (-1,0.5) --  (-0.5,0.25) node[above] {$\phi$}  -- (-0.089,0.047);
        \draw (-1,-0.5) node[left] {$~$} --  (-0.5,-0.25) node[below] {$\phi$}  -- (-0.089,-0.047);
        \draw (0.1,0) -- (0.5,0) node[above] {$\sigma$} -- (1,0);
    \end{tikzpicture} 
    } 
    &\quad\to\quad \int_x \frac{\tilde{g}}{2a}\, (...) \\
    \parbox{25mm}{
    \begin{tikzpicture}
        \draw (0,0) node[below] {$x$} circle (0.1);
        \fill[pattern=north east lines] (0,0) circle (0.1);
        \draw (-1,0.5) --  (-0.5,0.25) node[above] {$\phi$}  -- (-0.089,0.047);
        \draw (-1,-0.5) --  (-0.5,-0.25) node[below] {$\phi$}  -- (-0.089,-0.047);
        \draw (-1,0) node[left] {$\phi$} -- (-0.1,0);
        \draw (0.1,0) -- (0.5,0) node[above] {$\sigma$} -- (1,0);
    \end{tikzpicture} 
    } 
    &\quad\to\quad \int_x \frac{\lambda}{6} \, (...) 
\end{align}
where the shaded vertices can be either a $-$ node or a $+$ node, which are already encoded in Eq.~\eqref{eq: Dnm components}. It should be noted that, the first vertex given in \eqref{eq: phi-sigma} exactly vanishes by following the EOM~\eqref{eq: EOM of sigma}
\begin{align}
    (\partial_{\eta_1}^2 + 2\mathcal{H}_1\partial_{\eta_1} - \nabla_{\bm{x}}^2 +\tilde{m}^2) G^>(x,y) = 0~,
\end{align}
suggesting vanishing contribution to our result. This vertex corresponds to a transition between $\phi$ and $\sigma$, who share the same origin in the total field theory, and thereby simply reduces to a propagator of $\phi$ owing to the momentum conservation.

From the definition Eq.~\eqref{eq: G> thermal}, we can directly get
\begin{align}
    G^>(x,y)-G^<(x,y) = \int\frac{\md^3\bm{k}}{(2\pi)^3} \,\me^{i\bm{k} \cdot\bm{r}}  \frac{\chi_{{k}} (\eta_1)\chi^*_{{k}}(\eta_2) - \chi^*_{{k}} (\eta_1)\chi_{{k}}(\eta_2)}{a(\eta_1)a(\eta_2)}~.
\end{align}
It is noticed that, to the tree level, the particle occupation number $n_k$ does not contribute to the memory kernel.
For later convenience, let us define two real functions
\begin{align}
    \label{eq: def P}
    P_{{k}}(\eta_1,\eta_2) = -i\left(\chi_{{k}} (\eta_1)\chi^*_{{k}}(\eta_2) - \chi^*_{{k}} (\eta_1)\chi_{{k}}(\eta_2) \right) = 2\, \mathrm{Im}\,(\chi_{{k}} (\eta_1)\chi^*_{{k}}(\eta_2))~,\\
    \label{eq: def Q}
    Q_{{k}}(\eta_1,\eta_2) = \chi_{{k}} (\eta_1)\chi^*_{{k}}(\eta_2) + \chi^*_{{k}} (\eta_1)\chi_{{k}}(\eta_2)= 2\, \mathrm{Re}\,(\chi_{{k}} (\eta_1)\chi^*_{{k}}(\eta_2))~,
\end{align}
Since $\chi_{ k}$ and $\chi_{ k}^*$ are two independent solutions to EOM~\eqref{eq: EOM of chi}, $P_{ k}$ and $Q_{ k}$ satisfy the same equation, whose boundary conditions are determined by the Wronskian of $\chi_{ k}$. For a given $\eta_1$, the equations read
\begin{align}
    \label{eq: EOM of P}
    \left( \partial_{\eta}^2 + k^2 + a^2m^2\right) P_{{k}}(\eta_1, \eta) &= 0~, && 
    P_{ k}(\eta_1,\eta_1)=0~,~ 
    \partial_{\eta}P_{ k}(\eta_1,\eta)|_{\eta=\eta_1}=1~. \\
    \label{eq: EOM of Q}
    \left( \partial_{\eta}^2 + k^2 + a^2m^2\right) Q_{{k}}(\eta_1, \eta) &= 0~,&  &
    Q_{ k}(\eta_1,\eta_1)=2|\chi_{ k}(\eta_1)|^2~, \nonumber\\ 
    &&&\partial_{\eta}Q_{ k}
    (\eta_1,\eta)|_{\eta=\eta_1}=\left.\partial_{\eta}|\chi_{ k}(\eta)|^2\right|_{\eta=\eta_1}~.
\end{align}
After linear combinations, we can write down two independent real solutions to the EOM,
\begin{align}
    Y_k(\eta) = \me^{\gamma\eta^2/2} \,_1F_1\left(\frac{\mu}{2};\frac{1}{2};-\gamma\eta^2\right)~,~
    Z_k(\eta) = \eta\, \me^{\gamma\eta^2/2} \,_1F_1\left(\frac{\mu+1}{2};\frac{3}{2};-\gamma\eta^2\right)~,
\end{align}
by which the imaginary and real components of $G^>$ can be expressed as
\begin{align}
    P_k(\eta_1,\eta_2) &= Y_k(\eta_1)Z_k(\eta_2) - Z_k(\eta_1)Y_k(\eta_2)~,\\
    Q_k(\eta_1,\eta_2) &= \frac{Y_k(\eta_1)Y_k(\eta_2)}{\Omega_k} +\Omega_k Z_k(\eta_1)Z_k(\eta_2)~,\\
    \Omega_k &= \sqrt{4\gamma} \, \frac{\Gamma\left((\mu+1)/2\right)} {\Gamma\left(\mu/2\right)}~.
\end{align}
Therefore, the $G^{>} - G^{<}$ term is expressed by a converged integral
\begin{align}
    G^>(x,y)-G^<(x,y) = i\int_{k_W}^\infty \frac{k^2\md k}{2\pi^2} \frac{\sin(kr)}{kr} \frac{P_k(\eta_1,\eta_2)}{a(\eta_1)a(\eta_2)}~.
\end{align}

In the small $k_W$ limit with $k_W\ll m_0$, the vertices depend on time alone, and thereby the spatial average applies only to the propagators, which is given by
\begin{align}
    -i\int_{\bm{r}}\big(G^>(x,y)-G^<(x,y)\big) =\int\md^3\bm{r} \int\frac{\md^3\bm{k}}{(2\pi)^3} \,\me^{i\bm{k}\cdot\bm{r}}\, \frac{ P_k(\eta_1,\eta_2)}{a(\eta_1)a(\eta_2)} \simeq \frac{ P_{k_W}(\eta_1,\eta_2)}{a(\eta_1)a(\eta_2)}
    \label{eq: G> dr}
\end{align}
Consider an early time limit $\eta\ll\eta_0$, the story can be much simplified owning to a well-behaved propagator~\eqref{eq: chi plane wave}, in which case Eq.~\eqref{eq: G> dr} one has
\begin{align}\label{eq: Im G>}
    -i\int_{\bm{r}}\big(G^>(x,y)-G^<(x,y)\big) = 2\,\mathrm{Im}\int_{\bm{r}}\frac{ \chi_{0} (\eta_1)\chi^*_{0}(\eta_2)}{a(\eta_1)a(\eta_2)} = -\frac{\sin(m_0\Delta\eta)}{a(\eta_1)a(\eta_2)m_0}~,
\end{align}
with abbreviation $\Delta\eta \equiv \eta_1-\eta_2$.

\subsubsection{1-loop}\label{subsec: 1-loop}
The $1$-loop level contribution corresponds to the following diagram,
\begin{align}
    \parbox{30mm}{\centering
    \begin{tikzpicture}
        \draw (-1,0) node[below] {$x$} circle (0.1);
        \fill (1,0)  node[below] {$y$} circle (0.1);
        \draw (-1+0.089,0.047) .. controls (-0.5,0.4) and (0.5,0.4) .. (1,0);
        \draw (-1+0.089,-0.047) .. controls (-0.5,-0.4) and (0.5,-0.4) .. (1,0);
        \draw (0,0.3) node[above] {$\sigma$};
        \draw (0,-0.3) node[below] {$\sigma$};
    \end{tikzpicture}
    } ~-~ 
    \parbox{30mm}{\centering
    \begin{tikzpicture}
        \fill (-1,0) node[below] {$x$} circle (0.1);
        \draw (1,0)  node[below] {$y$} circle (0.1);
        \draw (-1,0) .. controls (-0.5,0.4) and (0.5,0.4) .. (1-0.089,0.047);
        \draw (-1,0) .. controls (-0.5,-0.4) and (0.5,-0.4) .. (1-0.089,-0.047);
        \draw (0,0.3) node[above] {$\sigma$};
        \draw (0,-0.3) node[below] {$\sigma$};
    \end{tikzpicture}
    }
    = G^>(x,y)^2-G^<(x,y)^2~,
\end{align}
with vertices given by
\begin{align}
    \parbox{25mm}{
    \begin{tikzpicture}
        \draw (0,0) node[below] {$x$} circle (0.1);
        \fill[pattern=north east lines] (0,0) circle (0.1);
        \draw (1,0.5) --  (0.5,0.25) node[above] {$\sigma$}  -- (0.089,0.047);
        \draw (1,-0.5)--  (0.5,-0.25) node[below] {$\sigma$}  -- (0.089,-0.047);
        \draw (-0.1,0) -- (-0.5,0) node[above] {$\phi$} -- (-1,0);
    \end{tikzpicture} 
    } 
    &\to\quad \int_x \frac{\tilde{g}}{2a}\, (...) \quad, \\
    \parbox{25mm}{
    \begin{tikzpicture}
        \draw (0,0) node[below] {$x$} circle (0.1);
        \fill[pattern=north east lines] (0,0) circle (0.1);
        \draw (-1,0.5) --  (-0.5,0.25) node[above] {$\phi$}  -- (-0.089,0.047);
        \draw (-1,-0.5) --  (-0.5,-0.25) node[below] {$\phi$}  -- (-0.089,-0.047);
        \draw (1,0.5) --  (0.5,0.25) node[above] {$\sigma$}  -- (0.089,0.047);
        \draw (1,-0.5)--  (0.5,-0.25) node[below] {$\sigma$}  -- (0.089,-0.047);
    \end{tikzpicture} 
    } 
    &\to\quad \int_x \frac{\lambda}{6}\, (...)
\end{align}
Different from the tree level, now $n_k$ will enter and enhance the contribution of $1$-loop result,
\begin{align}
    &G^>(x,y)^2-G^<(x,y)^2 = \left(G^>(x,y)-G^<(x,y)\right) \left(G^>(x,y) + G^<(x,y)\right) \nonumber\\
    =&\, i\int\frac{\md^3\bm{k}}{(2\pi)^3} \,\me^{i\bm{k}\cdot\bm{r}}\, \frac{P_k(\eta_1,\eta_2)}{a(\eta_1)a(\eta_2)} \times
    \int\frac{\md^3\bm{p}}{(2\pi)^3}  \,\me^{i\bm{p}\cdot\bm{r}}\, (1+2n_p)\frac{Q_p(\eta_1,\eta_2)}{a(\eta_1)a(\eta_2)} \nonumber\\
    \equiv &\, (\mathrm{tree~level~result}) \times (\mathrm{correction})
    \label{eq: G>2 - G<2}
\end{align}
Here again we consider the small $k_W$ limit. After spatial average, the two integrals convolved into one integral,
\begin{align}
    \label{eq: im of G>2}
    -i\int_{\bm{r}} \left(G^>(x,y)^2-G^<(x,y)^2\right)\simeq \int\frac{\md^3\bm{k}}{(2\pi)^3}(1+2n_k) \frac{P_k(\eta_1,\eta_2) Q_k(\eta_1,\eta_2)}{a(\eta_1)^2a(\eta_2)^2}~.
\end{align}
In the early time limit $\eta\ll\eta_0$, the $P_k$ and $Q_k$ functions can be replaced by its plane-wave counterpart,
\begin{align}
    & P_k(\eta_1,\eta_2) \simeq -\frac{\sin(E_k\Delta\eta)}{E_k}~, \quad
    Q_k(\eta_1,\eta_2) \simeq \frac{\cos(E_k\Delta\eta)}{E_k}~, \\
    &P_k(\eta_1,\eta_2)Q_k(\eta_1,\eta_2) \simeq \frac{1}{2E_k^2}\mathrm{Im}\,\me^{-i2E_k\Delta\eta}~.
\end{align}
Then we can rewrite Eq.~\eqref{eq: im of G>2} as
\begin{align}
    -i\int_{\bm{r}} \left(G^>(x,y)^2-G^<(x,y)^2\right) \simeq \int_0^\infty\frac{k^2\md k}{2\pi^2} (1+2n_k) \,\frac{1}{2}\mathrm{Im}\frac{\me^{-i2E_k\Delta\eta}}{a(\eta_1)^2a(\eta_2)^2E_k^2} \equiv I_\mathrm{vac} + I_\mathrm{th}~,
\end{align}
where $I_\mathrm{vac}$ and $I_\mathrm{th}$ correspond to the contributions associated with the two terms $1$ and $2n_k$, respectively. The integrand of the thermal term $I_\mathrm{th}$ is proportional to an exponentially decay occupation number $n_k$ at large $k$ to ensure its convergence. By contrast, the vacuum contribution term $I_\mathrm{vac}$ is not subject to any suppression and therefore need to be treated more carefully. 

We first compute $I_\mathrm{vac}$. Making the change of variable $u=E_k/m_0$ yields
\begin{align}\label{eq: I1 def}
    I_\mathrm{vac} &= \frac{1}{2a(\eta_1)^2a(\eta_2)^2} \mathrm{Im}\int_0^\infty\frac{\md k}{2\pi^2} \frac{k^2}{E_k^2} \,\me^{-i2E_k\Delta\eta} \nonumber\\ &= \frac{m_0}{4\pi^2a(\eta_1)^2a(\eta_2)^2} \mathrm{Im}\int_1^\infty \md u \, \frac{\sqrt{u^2-1}}{u} \me^{izu}~,
\end{align}
where $z = -2m_0\Delta\eta$. Without losing generality, we first assume $\Delta \eta<0$ and $z>0$ here, as the other sign branch can be obtained by taking a complex conjugate of the integrand in \eqref{eq: I1 def}. Let us focus on the integral and define
\begin{align}
    \label{eq: def Fz}
    F(z) \equiv\int_1^\infty \md u \, \frac{\sqrt{u^2-1}}{u} \me^{izu}~,\quad F'(z)=i \int_1^\infty \md u \, \sqrt{u^2-1} \, \me^{izu}~.
\end{align}
Consider the following integral,
\begin{align}
    B(s)=\int_1^\infty \md u \, \sqrt{u^2-1} \, \me^{-su} =\frac{K_1(s)}{s}~,\quad \mathrm{for} ~s>0~,
\end{align}
where $K_\nu(s)$ is the modified (hyperbolic) Bessel function of the second kind, which is equivalent to the analytical continuation of the Hankel function of the first kind,
\begin{align}\label{eq: K to H}
    K_\nu(s)=\frac{i\pi}{2}\me^{i\pi\nu/2} H_\nu^{(1)}(is) 
    \quad\Rightarrow\quad K_\nu(-iz) = \frac{i\pi}{2}\me^{i\pi\nu/2} H_\nu^{(1)}(z)~.
\end{align}
Then one has
\begin{align}\label{eq: Fprime}
    F'(z) = i \lim_{\epsilon\to 0^+} B(\epsilon - iz) = i\lim_{\epsilon\to 0^+} \frac{K_1(\epsilon - iz)}{\epsilon + iz} = \frac{\pi}{2}\frac{H_1^{(1)}(z)}{z}~,
\end{align}
where $\epsilon$ is required to be greater than $0$ to ensure convergence. Set a boundary value $F(\infty)=0$, one can then integrate $F'$ and arrive at
\begin{align}
    \label{eq: F as H}
    F(z) &= \int^z_\infty F'(u)\md u = -\frac{\pi}{2}\int_z^\infty\frac{H_1^{(1)}(u)}{u}\md u ~.
\end{align}
We are mainly interested in the imaginary part of the integral, which gives
\begin{align}\label{eq: Im G>2}
    I_\mathrm{vac} = -\, \frac{m_0}{8\pi a(\eta_1)^2a(\eta_2)^2} \mathrm{Im}\int_{|z|}^\infty \frac{H_1^{(1)}(u)}{u}\md u~.
\end{align}
For the case $z<0$, we take an absolute value of $z$ in the lower limit of the integral by using the fact $\mathrm{Im}\,H_1^{(1)}(-z) = -\mathrm{Im}\,H_1^{(1)}(z)$ for a real $z$. In fact, we can expand the Hankel function in terms of Bessel functions $H_\nu^{(1)}(z)=J_\nu(z) + i Y_\nu(z)$, yielding
\begin{align}
    I_\mathrm{vac} = \mathrm{sgn}(\Delta\eta)\, \frac{m_0}{8\pi a(\eta_1)^2a(\eta_2)^2} \int_{2m_0|\Delta\eta|}^\infty \frac{Y_1(u)}{u}\md u~.
\end{align}

Next we compute $I_\mathrm{th}$, which is naturally converged,
\begin{align}
    I_\mathrm{th} = \frac{1}{a(\eta_1)^2a(\eta_2)^2} \mathrm{Im}\int_0^\infty\frac{\md k}{2\pi^2} \frac{k^2}{E_k^2} \frac{1}{\me^{E_k/E_i}-1}\,\me^{-i2E_k\Delta\eta}~,
\end{align}
where $E_i \equiv a_iT_i$. Then we expand $n_k$ into a series
\begin{align}
    n_k = \frac{1}{\me^{E_k/E_i}-1} = \sum_{n=1}^\infty \me^{-nE_k/E_i}~.
\end{align}
Using the same change of variable $u=E_k/m_0$ and definition for $F(z)$~\eqref{eq: def Fz} gives
\begin{align}
    I_\mathrm{th} &= \frac{m_0}{2\pi^2 a(\eta_1)^2a(\eta_2)^2} \, \mathrm{Im}\sum_{n=1}^{\infty} \,F\left( -2m_0\Delta\eta + i n\frac{m_0}{E_i}\right) \nonumber\\
    &= \frac{m_0}{4\pi^2 a(\eta_1)^2a(\eta_2)^2} \int_{0}^\infty \md v \, \frac{Y_1(v)}{v} \sum_{n=1}^\infty \mathcal{S}\left(v,2m_0\Delta\eta, n\frac{m_0}{E_i}\right) \equiv\sum_{n=1}^\infty I_{\mathrm{th}}^{(n)}~,
    \label{eq: I th}
\end{align}
where $\mathcal{S}$ is defined as
\begin{align}
    \mathcal{S}(v,x,a_n) = \arctan\frac{v-x}{a_n} - \arctan\frac{v+x}{a_n} + 2\arctan\frac{x}{a_n}~.
\end{align}
The detailed derivation of Eq.~\eqref{eq: I th} is complicated but straightforward, where the mathematical tricks used are provided in Appendix.~\ref{app: integrals F}. We can make an order-of-magnitude estimation in two limits to understand this result. For the limit $nm_0/E_i\gg 1$, $\mathcal{S}$ (as well as $n_k$) decays exponentially, the thermal contribution goes to zero. In the opposite regime, we consider the limit $a_n=nm_0/E_i\ll1$ for a fixed summation index $n$,
\begin{align}
    \lim_{a_n\to0}\mathcal{S}\left(v,2m_0\Delta\eta, a\right) = \mathrm{sgn}(\Delta\eta) \pi \, \Theta\left(v-2m_0|\Delta\eta|\right)~.
\end{align}
Consequently, the corresponding contribution can be written as
\begin{align}
    I_\mathrm{th}^{(n)} \simeq \mathrm{sgn}(\Delta\eta)\, \frac{m_0}{4\pi a(\eta_1)^2a(\eta_2)^2} \int_{2m_0|\Delta\eta|}^\infty \frac{Y_1(u)}{u}\md u = 2I_\mathrm{vac}~,
\end{align}
that is, each term in the thermal expansion approaches twice of the vacuum contribution. As a rough estimation, we introduce an effective cutoff $N$, below which the thermal contribution can be approximated as twice of the vacuum for each excitation state, while above which it can be neglected. Therefore, the total contribution can be estimated as $I_\mathrm{vac} + I_\mathrm{th} \simeq (1+2N)I_\mathrm{vac}$.
This picture can be understood by recalling the $1+2n_k$ structure appearing in Eq.~\eqref{eq: im of G>2}, where the unity corresponds to vacuum and $2n_k$ accounts for thermal excitations. Our approximation effectively replaces the $k$-dependent occupation number $n_k$ with an effective number $N\simeq E_i/m_0$, which is determined only by the initial data. Although this sharp-cutoff approximation is only qualitative, it captures the thermal enhancement of the vacuum contribution.

The $G^{>,3} - G^{<,3}$ term in $D_{11}$ corresponds to the following sunrise diagram,
\begin{align}
    \parbox{30mm}{\centering
    \begin{tikzpicture}
        \draw (-1,0) node[below] {$x$} circle (0.1);
        \fill (1,0)  node[below] {$y$} circle (0.1);
        \draw (-1+0.089,0.047) .. controls (-0.5,0.5) and (0.5,0.5) .. (1,0);
        \draw (-1+0.089,-0.047) .. controls (-0.5,-0.5) and (0.5,-0.5) .. (1,0);
        \draw (-0.9,0) -- (0,0) -- (0.9,0);
        \draw (0,0.3) node[above] {$\sigma$};
        \draw (0,-0.3) node[below] {$\sigma$};
        \draw (0,-0.15) node {$\sigma$};
    \end{tikzpicture}
    } ~-~ 
    \parbox{30mm}{\centering
    \begin{tikzpicture}
        \fill (-1,0) node[below] {$x$} circle (0.1);
        \draw (1,0)  node[below] {$y$} circle (0.1);
        \draw (-1,0) .. controls (-0.5,0.5) and (0.5,0.5) .. (1-0.089,0.047);
        \draw (-1,0) .. controls (-0.5,-0.5) and (0.5,-0.5) .. (1-0.089,-0.047);
        \draw (-0.9,0) -- (0,0) -- (0.9,0);
        \draw (0,0.3) node[above] {$\sigma$};
        \draw (0,-0.3) node[below] {$\sigma$};
        \draw (0,-0.15) node {$\sigma$};
    \end{tikzpicture}
    }
    = G^>(x,y)^3-G^<(x,y)^3~,
\end{align}
which in principle can be calculated following similar procedures.
Since this term corresponds to the $2$-loop level contribution, we leave its detailed investigation for future work.

\subsubsection{Markovian approximation}

We name $\mathbf{D}$ as a memory kernel, since its temporal nonlocality implies history dependence of the dynamics, suggesting that the system is non-Markovian in the generalized-Langevin sense.
Consequently, the motion of $\phi_c$ at present time $\eta$ gains dependence on its past trajectory $\phi(\eta')$ with $\eta'<\eta$. If further assuming the random-walk time scale $T_\mathrm{random}$ for $\phi_c$ is much shorter than the variation time scale of the memory kernel $T_\mathrm{memory}$, one can apply the Markov approximation by Taylor expanding the field around present time. Keeping to the leading order $\phi_c(\eta') \simeq \phi_c(\eta) - \Delta\eta \partial_\eta\phi_c(\eta)$ helps us to extract $\phi_c$ and the corresponding derivatives out of the integral, leaving the memory kernel independent of the field history,
\begin{align}
    &\int_0^{\eta}\md\eta' \, a(\eta')^4 \, \bar{\mathbf{D}}(\eta,\eta') \mathbf{J}_c^{\mathrm{lin}}(\eta') \simeq \nonumber\\
    & \left(\int_0^{\eta}\md\eta' \, a(\eta')^4 \, \bar{\mathbf{D}}(\eta,\eta') \right) \mathbf{J}_c^{\mathrm{lin}}(\eta) - \left(\int_0^{\eta}\md\eta' \, a(\eta')^4 \, \bar{\mathbf{D}}(\eta,\eta')\Delta\eta \right) \frac{\partial 
    \mathbf{J}_c^{\mathrm{lin}}(\eta)}{\partial\eta}~,
\end{align}
which is composed of a purely drifting term (proportional to polynomials of $\phi_c$) and a dissipation term (proportional to $\phi_c'$).
From Eq.~\eqref{eq: Im G>} and \eqref{eq: Im G>2}, one can see that the approximation does not hold for heavy field, as the typical oscillating period of the memory kernel is inversely proportional to the mass $T_\mathrm{memory}\simeq 2\pi/m_0$. However, this does not guarantee that the approximation always holds for light fields, since the memory kernels still retain time dependence in the limit the limit $m_0\to 0$.

\subsection{Noise term}\label{subsec: noise}

The noises are characterised by the correlation functions defined in \eqref{eq: Ci def}, with the corresponding $K_i$ directly  identified from \eqref{eq: Seff final}. Here we focus on only the leading-order Gaussian noise contribution for simplicity, where the statistical properties are totally determined by the real part of two-point correlation function of $\sigma$. The higher-order noises corresponding to non-Gaussianity can be explored following a similar procedure. Similar to the discussions on the memory kernel, the central structure of two-point correlation function of noise now becomes $G^{>,n} + G^{<,n}$, corresponding to the real part of $G^{>,n}$, i.e. $ \mathbf{N}(\eta_1,\eta_2;\bm{r}) \supset \mathrm{Re}\, G^>(\eta_1,\eta_2;\bm{r})^n$. The simplest case with $n=1$ case is just the ``correction'' factor in Eq.~\eqref{eq: G>2 - G<2}, receiving contributions both from vacuum and thermal excitation.

In the small $k_W$ limit, a coarse-grained length-scale $L$ can be introduced to define the spatial-averaged noise $\bar{\xi}$. Here we discuss the correlation functions in two limits, and then provide a general expression for arbitrary $L$. For the large scale limit, the correlation functions are highly suppressed by the spatial volume. From the definition, it is easy to see that, for a sufficiently large $\mathcal{V}$, the $n$-point correlation functions are suppressed by a factor of $\mathcal{V}^{n-1}$. For example, the two-point correlator is given by
\begin{align}\label{eq: xibar two-p}
    \langle\bar{\xi}(\eta_1)\bar{\xi}(\eta_2)\rangle &= \frac{1}{\mathcal{V}^2} \mathbf{J}_\Delta^{\mathrm{lin}}(\eta_1)^\top \iint_\mathcal{V} \md^3\bm{x}\md^3\bm{y}  \, \mathbf{N}(\eta_1,\bm{x};\eta_2,\bm{y}) \mathbf{J}_\Delta^\mathrm{lin}(\eta_2) \nonumber \\
    & \simeq \frac{1}{\mathcal{V}^2} \mathbf{J}_\Delta^{\mathrm{lin}}(\eta_1)^\top \, \mathcal{V} \, \bar{\mathbf{N}}(\eta_1,\eta_2) \,\mathbf{J}_\Delta^\mathrm{lin}(\eta_2) \propto \mathcal{V}^{-1}
\end{align}
For an extended $K_n$, the suppression factor could be less stronger. If the environment field $\sigma$ were massless, an IR divergence appear in the propagators, yielding a divergence in $\bar{\mathbf{N}}$. This divergence would cancel the overall $\mathcal{V}^{-1}$ factor in \eqref{eq: xibar two-p} yielding a non-vanishing and volume-independent two-point correlation function, consistent with the long-range correlations carried by massless modes. However, in the case we study, $\sigma$ must be massive, as it shares the same origin with $\phi$, which requires a strictly positive mass to ensure the stability of the false vacuum $\Phi=0$ in the early time $\eta\to 0$. Consequently, the suppression is inevitable. 

Physically speaking, the suppression arises from a coarse-graining the system. Increasing the spatial volume incorporates more small-scale modes, whose contributions partially cancel under volume averaging, thereby reducing spatial correlations on average. The decorrelation of noise on large scales can be understood intuitively through an analogy with waves on the ocean surface. On small spatial scales, the surface is highly irregular and the local height exhibits strong fluctuations due to turbulent wave motion. However, when the surface is coarse-grained over length scales much larger length scale, these fluctuations tend to cancel through spatial averaging. As a result, the large-scale mean surface level remains essentially unchanged. Under the limit $L\to \infty$, all correlations vanish and the EOM becomes a noise-free equation. By setting initial condition $\phi_c(\eta_\mathrm{ini})=0,~\dot{\phi}_c(\eta_\mathrm{ini})=0$, the solution is a trivial solution $\phi_c(\eta)$ which is trapped in the local minimum. This result is expected, since the stochastic noise is turned off in this limit. In the absence of quantum fluctuations, the system reduces to purely classical evolution, and the classical trajectory therefore remains trapped in the false vacuum.

For the small scale limit with $L\to 0$, the noises only exhibit local correlations in space,
\begin{align}
    \langle\bar{\xi}(\eta_1)\bar{\xi}(\eta_2)\rangle =  \mathbf{J}_\Delta^{\mathrm{lin}}(\eta_1)^\top \, \mathbf{N}(\eta_1,\bm{0};\eta_2,\bm{0}) \mathbf{J}_\Delta^\mathrm{lin}(\eta_2)~.
\end{align}
By definition the central structure is given by
\begin{align}\label{eq: G> r0}
    G^>+G^<=2\,\mathrm{Re}\,G^>(\eta_1,\eta_2;\bm{0}) =2\int\frac{\md^3\bm{k}}{(2\pi)^3} (1+2n_k)\frac{Q_k(\eta_1,\eta_2)}{a(\eta_1)a(\eta_2)}~.
\end{align}
In the early time limit it becomes
\begin{align}\label{eq: G> r0 early}
    2\,\mathrm{Re}\,G^>(\eta_1,\eta_2;\bm{0}) &\simeq \mathrm{Re}\int\frac{\md^3\bm{k}}{(2\pi)^3} (1+2n_k)\frac{\me^{-iE_k\Delta\eta}}{E_k} 
    \\
    & = \frac{m_0^2}{4\pi^2}\,\mathrm{Re}\int_1^\infty \md u \, \sqrt{u^2-1} \, \me^{-iu m_0\Delta \eta} \left(1+2\sum_{n=1}^\infty\me^{-a_n u}\right) \nonumber\\
    &\equiv J_\mathrm{vac,0} + J_\mathrm{th,0}~.
\end{align}
where $a_n = nm_0/E_i$. Let us take a closer look at the vacuum contribution.
From Eq.~\eqref{eq: Fprime} one can get
\begin{align}
    \int_1^\infty \md u \, \sqrt{u^2-1} \, \me^{-iu m_0\Delta \eta} ~\longrightarrow~ i \frac{\pi}{2} \frac{H_1^{(1)}(- m_0\Delta \eta)}{m_0\Delta \eta}~.
\end{align}
Again using the fact $\mathrm{Im}\,H_1^{(1)}(-z) = -\mathrm{Im}\,H_1^{(1)}(z)$ yields
\begin{align}
    J_\mathrm{vac,0} =  \frac{m_0^2}{4\pi^2} \cdot \frac{\pi}{2} \frac{\mathrm{Im}\, H_1^{(1)}(m_0\Delta \eta)}{m_0\Delta \eta} = \frac{m_0}{8\pi\Delta\eta} Y_1(m_0\Delta \eta)~.
\end{align}
It should be noted that the above result becomes divergent in the limit $\Delta\eta \to 0$, which arises from contributions all $k$ modes. In the absence of a UV-cutoff, the theory receives contributions from arbitrarily high-$k$ modes, causing local correlators to develop the short-distance singularity in the limit $x \to y$.

For a finite cutoff $L$, we have to compute the following spatial average,
\begin{align}\label{eq: G> average}
    \mathcal{J} &\equiv \frac{1}{\mathcal{V}^2} \iint \md^3\bm{x} \md^3 \bm{y} \,\mathrm{Re}\,G^>(x,y) \nonumber\\
    &= \frac{1}{\mathcal{V}^2}\int\frac{\md^3\bm{k}}{(2\pi)^3} 
    (1+2n_k) \frac{Q_k(\eta_1,\eta_2)}{a(\eta_1)a(\eta_2)}
    \int\md^3\bm{x} \,\me^{i\bm{k}\cdot\bm{x}} \int\md^3\bm{y} \,\me^{-i\bm{k}\cdot\bm{y}}~.
\end{align}
The spatial integrals are given by
\begin{align}
    \int\md^3\bm{x} \,\me^{i\bm{k}\cdot\bm{x}} &=2\pi \int_0^L \md r~r^2\int_{-1}^1\md\mu \, \me^{ik r\mu}= 4\pi \int_0^L \md r \, r^2j_0(kr) = 4\pi \frac{L^2}{k} j_1(kL)~,
\end{align}
where $j_\nu(x)$ is the $\nu$-th order spherical Bessel function. The spatial volume is related to the cutoff as $\mathcal{V}=4\pi L^3/3$, and then Eq.~\eqref{eq: G> average} can be recast to
\begin{align}\label{eq: G> dxdy}
    \mathcal{J} = \frac{9}{4\pi^2 L^2} \int_0^\infty \md k \, j_1(kL)^2 \, (1+2n_k) \frac{Q_k(\eta_1,\eta_2)}{a(\eta_1)a(\eta_2)}~.
\end{align}
As a consistency check, Eq.~\eqref{eq: G> dxdy} reduces back to \eqref{eq: G> r0} by noticing the fact that
\begin{align}
    \lim_{L\to 0} \frac{j_1(kL)}{kL} = \frac{1}{3}~.
\end{align}
As $j_1(kL)$ vanishes for $k\to \infty$, it suppresses the contributions from high-$k$ modes, serving as a UV-cutoff of the theory. Here we evaluate the integrals in \eqref{eq: G> dxdy} explicitly. Firstly, project $j_1$ in terms of $j_0$ as
\begin{align}
    j_1^2(z) = \frac{1}{2}\int_{-1}^1 \md \mu ~\mu j_0\left(z\sqrt{2(1-\mu)}\right)~.
\end{align}
Define abbreviation $R=L\sqrt{2(1-\mu)}$, and substitute the projection above into the integral,
\begin{align}
    \mathcal{J} = \frac{9}{8\pi^2L^2}\int_{-1}^1 \md\mu ~ \mu\int_0^\infty \md k \,j_0(kR) (1+2n_k) \frac{Q_k(\eta_1,\eta_2)}{a(\eta_1)a(\eta_2)}~.
\end{align}
In the early time limit, we can consider the following integral,
\begin{align} 
    \tilde{\mathcal{J}} &= \frac{9}{8\pi^2L^2}\int_{-1}^1 \md\mu ~ \mu\int_0^\infty \md k \,j_0(kR) \frac{\me^{-iE_k\Delta\eta} }{E_k} \left(1+2\sum_{n=1}^\infty\me^{-a_n u}\right) \nonumber\\
    &=\frac{9}{8\pi^2L^2}\int_{-1}^1 \md\mu ~ \mu
    \int_0^R \frac{\md r}{R}
    \int_0^\infty \md k \,\cos(kr) \frac{\me^{-iE_k\Delta\eta} }{E_k} \left(1+2\sum_{n=1}^\infty\me^{-a_n u}\right)~,\\
    \mathcal{J} &= \mathrm{Re}\,\tilde{\mathcal{J}} \equiv J_\mathrm{vac} + \sum_{n=1}^\infty J_\mathrm{th}^{(n)}~,
    \label{eq: J vac and th}
\end{align}
where we have used the fact that 
\begin{align}
    j_0(kR) = \frac{\sin(kR)}{kR} = \frac{1}{R} \int_0^R \cos (k r)\, \md r~.
\end{align}
The integral over $k$ can be performed by the formula provided in Appendix.~\ref{app: integrals Bessel},
\begin{align}
    \int_0^\infty \md k \, \frac{\cos kr}{E_k} \,\me^{-iE_k(\Delta\eta - ia_n)} = K_0\left(m_0\sqrt{r^2-(\Delta\eta-i a_n)^2}\right)~.
\end{align}
This result is well defined as long as $a_n>0$. For the vacuum case with $a_0=0$, one can make a change of variable $\Delta\eta\to\Delta\eta-i\epsilon$ and take the $\epsilon\to 0^+$ limit.
Under the the change of variable $\md\mu = -2R\md R/L^2$, the integral can be reduced to a $2D$ integral on the $R-r$ plane,
\begin{align}
    \mathcal{I}_n &\equiv \int_0^{2L} \frac{\md R}{R} \left(1-\frac{R^2}{2L^2}\right) \int_0^R \md r\, K_0\left(m_0\sqrt{r^2-(\Delta\eta-i a_n)^2}\right) \nonumber\\
    &= \int_0^{2L} \md r \left(\frac{1}{3L} - \frac{r}{2L^2} + \frac{r^3}{12L^4}\right) \, K_0\left(m_0\sqrt{r^2-(\Delta\eta-i a_n)^2}\right)~\nonumber\\
    &\equiv \int_0^{2L} \md r \, W_L(r) K_0\left(m_0\sqrt{r^2-(\Delta\eta-i a_n)^2}\right)
    \label{eq: cur I def}
\end{align}
where in the second line we have performed the integral over $R$ by interchanging the order of integration as follows,
\begin{align}
    \int_0^{2L}\md R \int_0^R \md r \, (\dots) = \int_0^{2L} \md r \int_r^{2L} \md R \, (\dots)~.
\end{align}
At last, the two contributions defined in Eq.~\eqref{eq: J vac and th} can be recast to
\begin{align}
    J_\mathrm{vac} &= \frac{9}{8\pi^2L^2} \mathrm{Re} \,\mathcal{I}_0~,\quad J_\mathrm{th}^{(n)} = \frac{9}{4\pi^2L^2} \mathrm{Re} \,\mathcal{I}_n~.
\end{align}

In general there is no simpler expression for the real part of $K_0$ with a complex argument. Nevertheless, the vacuum contribution case with $a_n=0$ can be further simplified. For the region $\Delta\eta < r < 2L$, $K_0$ is purely real and the integral is converged. For the region $0<r<\Delta\eta$, using \eqref{eq: K to H} and decomposition of Hankel function, one has
\begin{align}
     K_0 \left(iz\right) = -\frac{\pi}{2} Y_0 \left(z\right) - i \frac{\pi}{2} J_0(z)~.
\end{align}
Therefore, the real part of the integral is evaluated as
\begin{align}
    \mathrm{Re}\,\mathcal{I}_0 = -\frac{\pi}{2}\int_0^{\Delta\eta} \md r\, W_L(r) Y_0\left(m_0\sqrt{\Delta\eta^2 - r^2}\right) + \int_{\Delta\eta}^{2L} \md r\, W_L(r) K_0\left(m_0\sqrt{r^2-\Delta\eta^2}\right)~.
\end{align}
Although there is a pole at $r=\Delta\eta$ in the integrands, the total integral is converged. Taking the limit $L\to \infty$, $\mathrm{Re}\,\mathcal{I}$ goes to zero as expected. With similar arguments in Sec.~\ref{subsec: 1-loop}, we can take an approximation $J_\mathrm{th}^{(n)}\simeq2J_\mathrm{vac}$ for $n\lesssim N=E_i/m_0$, and $J_\mathrm{th}^{(n)}\simeq0$ for $n>N$, and the total contribution can be estimated as $\mathcal{J}=(1+2N)J_\mathrm{vac}$. The results for the higher-order structure $G^{>,n}$ can in principle be obtained by taking integrals with respect to $\Delta\eta$ in Eq.~\eqref{eq: cur I def}.

To close this section, we need to specify the comoving length scale $L$. A natural choice in an expanding universe is to identify it as the Hubble radius $L\simeq 1/\mathcal{H}$. As there are two time scales $\eta_1$ and $\eta_2$ (we have assumed $\eta_1>\eta_2$ without losing generality), $L$ should cover contributions on both scales, thus it matches the larger Hubble radius $L=1/\mathcal{H}_1 = \eta_1$. Physically, one may also choose it to be slightly larger than the vacuum bubble size at the nucleation time. For an order-of-magnitude estimation, the relevant scale lies between the average bubble radius and the Hubble radius.


\section{Conclusions and Discussions}\label{sec: conclusions}

In this work, we systematically constructed the Langevin description of cosmological vacuum phase transitions by Schwinger-Keldysh formalism. We split the total phase transition field $\Phi$ into the long-wavelength field $\phi$ and the short-wavelength modes $\sigma$, the latter of which is treated as environment fluctuations. We use a polynomial potential an explicit example and calculate the memory kernels and correlation function for the noises in the EOM. Nevertheless, the method can in general be applied to arbitrary potentials. 

By tracing out as environmental degrees of freedom $\sigma$, we arrive at a Langevin-type EOM for $\phi$. It is found that only the combination of linear terms in the current $\mathbf{J}_\Delta^\mathrm{lin}$ and $\mathbf{J}_c^\mathrm{lin}$ contributes to the deterministic memory terms in the EOM, with a kernel $\mathbf{D}$ made up of the imaginary part of propagators of $\sigma$. 
The bilinear term of $\mathbf{J}_\Delta^\mathrm{lin}$ is translated to the two-point correlation function of the noise $\xi$, while any other terms containing non-linear current $\mathbf{J}_\Delta^\mathrm{nl}$ or $\mathbf{J}_c^\mathrm{nl}$ would contribute to higher-order correlation functions for the noise. Since a phase transition field $\Phi$ is associated with a potential $V$ featuring at least two local minima (i.e., two vacua), the potential necessarily contains higher-order vertices beyond the quadratic mass term, thereby inevitably leading to a non-Gaussian noise in the Langevin equation, together with a non-Markov properties to the memory kernel. In addition to the non-Gaussianity, the non-vanishing two-point correlation in time also indicates a coloured noise instead of a simple white noise.

The stochastic EOM \eqref{eq: Langevin EOM} provides a semi-classical tool to simulate the vacuum bubble nucleation and expansion simultaneously. However, the presence of non-Markov effect in the memory kernel and the non-trivial temporal correlations of the noises may bring challenges to such a numerical simulation, which is left for future studies.
As a next step, it would be valuable to study the Fokker–Planck equation corresponding to Eq.~\eqref{eq: Langevin EOM}. Such a formulation would characterize the evolution of the probability distribution of the system. In particular, deriving its explicit form and obtaining corresponding numerical solutions may help clarify the long-time properties of the system and extract the vacuum decay rate, which is left as future studies. Nevertheless, it should be reminded that non-Gaussian colored noise generally leads to non-local memory effects and higher-order noise cumulants, rather than a closed, time-local Fokker–Planck equation~\cite{Athanassoulis:2019,Athanassoulis:2026}, thereby posing substantial challenges for both analytical estimation and numerical implementation.

\acknowledgments
The authors thank Shao-Jiang Wang, Xiao-Quan Ye, Rui-Chen Liu, Hao Sun, Zeyu Li, and Hai-Long Fu, for helpful discussions. The authors are grateful for the workshop at the Yukawa Institute for Theoretical Physics (YITP-T-26-05) for giving a chance to deepen their ideas. Z.-Y. Yuwen is supported by an appointment to the Young Scientist Training (YST) program at the APCTP through the Science and Technology Promotion Fund and Lottery Fund of the Korean Government. This was also supported by the Korean Local Governments-Gyeongsangbuk-do Province and Pohang City.

\appendix

\section{Integral Involving $F$ with complex arguments}\label{app: integrals F}

In this appendix, we derived the simplified expression for $F$ function with complex arguments. In particular, we only care about the imaginary part of $F$,
\begin{align}
    \mathrm{Im}\, F(-x + ia) &= \mathrm{Im} \int_1^\infty\md u \,\frac{\sqrt{u^2-1}}{u} \me^{-ixu-au} \nonumber\\ &= -\int_1^\infty\md u \,\frac{\sqrt{u^2-1}}{u} \me^{-au}\sin(xu)
    \label{eq: F -x+ia}
\end{align}
For $u>0$, the Fourier transform of exponential function is given by
\begin{align}
    \me^{-ixu-au} = \int_{-\infty}^{\infty} \md v \, \frac{a}{\pi\left((x-v)^2+a^2\right)} \,\me^{ivu}~.
\end{align}
The imaginary part can be represented in terms of a Fourier sine transform,
\begin{align}
    \me^{-au}\sin(xu) &= \int_0^{\infty} \md v \, \frac{a}{\pi} \left( \frac{1}{(x-v)^2+a^2} - \frac{1}{(x+v)^2+a^2}\right) \sin(uv) \nonumber\\ &\equiv \int_0^\infty\md v \,\mathcal{T}(v,x,a) \sin(uv)~.
\end{align}
Substituting this transform back into Eq.~\eqref{eq: F -x+ia} leads to
\begin{align}
    \mathrm{Im}\, F(-x + ia) &= -\int_0^\infty\md v\,\mathcal{T}(v,x,a) \int_1^\infty\md u \,\frac{\sqrt{u^2-1}}{u} \sin(vu) \nonumber\\
    &= \int_0^\infty\md v\,\mathcal{T}(v,x,a) \,\mathrm{Im}\,F(-v)~,
\end{align}
According to Eq.~\eqref{eq: F as H}, the imaginary part of $F$ function with a real positive argument $v$ can be expressed as
\begin{align}
    \mathrm{Im}\, F(-v) = \frac{\pi}{2} \int_{v}^\infty \frac{Y_1(u)}{u} \md u~,\quad v>0~. 
\end{align}
Define an auxiliary function $\mathcal{S}$ satisfying $\partial_v\mathcal{S}(v,x,a) = \pi \, \mathcal{T}(v,x,a)$, then we can perform the integral by parts to get
\begin{align}
    \mathrm{Im}\, F(-x + ia) = \mathcal{S}(v,x,a)\mathrm{Im}\, F(-v)\big|_0^\infty + \frac{1}{2} \int_0^\infty \md v \, \frac{Y_1(v)}{v}\mathcal{S}(v,x,a)~.
\end{align}
As $v\to\infty$, the the boundary term at infinity naturally goes to zero. We can choose a vanishing initial value for $\mathcal{S}$ such that the other boundary term at origin also gives no contribution, then
\begin{align}
    \mathcal{S}(v,x,a) &= \int_0^v \pi \, \mathcal{T}(v,x,a) = \arctan\frac{v-x}{a} - \arctan\frac{v+x}{a} + 2\arctan\frac{x}{a}~.
\end{align}

\section{Integral Involving Modified Bessel Functions}\label{app: integrals Bessel}

In this appendix, we provide the integral associated with modified Bessel functions used in the main text. We consider the following integral,
\begin{align}
    I \equiv \int_0^\infty \md k \, \frac{\cos kr}{E_k} \,\me^{-iE_k T}~,
\end{align}
where $T=t-i\tau$ satisfying $\tau>0$. First of all, consider a real number $a>0$ and an integral
\begin{align}\label{eq: I tilde}
    \tilde{I} = \int_0^\infty \md k \, \frac{\cos kr}{E_k} \,\me^{-E_k a}.
\end{align}
Substituting the Schwinger parametrization
\begin{align}
    \frac{\me^{-E_k a}}{E_k} = \frac{1}{\sqrt{\pi}}\int_0^\infty\frac{\md s}{\sqrt{s}} \exp\left(-E_k^2 s - \frac{a^2}{4s}\right)
\end{align}
back into \eqref{eq: I tilde} leads to
\begin{align}
    \tilde{I} &= \frac{1}{\sqrt{\pi}} \int _0^\infty\frac{\md s}{\sqrt{s}}\exp\left(-m^2 s - \frac{a^2}{4s}\right) \int_0^\infty\md k \, \cos (kr) \,\me^{-k^2s} \nonumber\\
    &= \frac{1}{2}\int _0^\infty\frac{\md s}{s} \exp\left(-m^2 s - \frac{a^2+r^2}{4s}\right)~,
\end{align}
which just matches one of the integral representations of $K_0$,
\begin{align}
    \tilde{I} = K_0\left(m\sqrt{r^2+a^2}\right)~.
\end{align}
This expression extends from a positive real $a$ to the entire half-plane $\mathrm{Re}\,a>0$ by analytic continuation. Then taking $a=iT = \tau + it$ directly gives
\begin{align}
    I = K_0\left(m\sqrt{r^2-T^2}\right)~.
\end{align}

\bibliographystyle{JHEP}
\bibliography{ref}

\end{document}